\documentclass[final,10pt]{iopart}
\usepackage{graphicx}
\begin{document}
\title[THz pulse accompanying laser wakes in plasma columns and channels]{Single-cycle THz signal accompanying laser wake in photoionized plasmas and plasma channels}

\author{S Y Kalmykov$^1$\footnote{Author to whom any correspondence should be addressed}, A Englesbe$^2$\footnote{Present address: Plasma Physics Division, Naval Research Laboratory, Washington, DC 20375, USA}, J Elle$^2$ and A Schmitt-Sody$^2$}
\address{$^1$ Leidos Innovations Center, Leidos Inc., 2109 Air Park Road SE, Ste 200, Albuquerque, NM 87106, USA}
\address{$^2$ High Power Electromagnetics Division, Air Force Research Laboratory, Kirtland Air Force Base, NM 87117, USA}
\ead{serge.kalmykov@leidos.com}

\begin{abstract}
Photoionization by a femtosecond, terawatt laser pulse generates a plasma column in a neutral ambient gas.\ Velocity of electrons, pushed by the laser ponderomotive force along the column surface, couples to the the radial density gradient at the column boundary, generating an azimuthally polarized THz rotational current (RC).\ The same mechanism produces the low-frequency RC in a leaky plasma channel.\ Applying external voltage to the channel induces a radially non-uniform electron flow (direct current) and a constant, azimuthally polarized magnetic field. Coupling them to the electron density perturbations adds two more terms to the RC. The surface RC in the plasma column supports a broadband, evanescent THz signal accompanying the wake. A few millimeters away from the column, rapid evanescence of high-frequency components turns this THz signal into a radially polarized, single-cycle pulse.
\end{abstract}

\submitto{J. Phys.: Conf. Series} \maketitle

\section{Introduction}
Multi-dimensional electron plasma oscillations \cite{Dawson_PR_1959} had been on the radar of advanced accelerator community for decades.\ A particular interest was expressed in electron plasma waves (EPWs), dubbed ``laser wakes'', driven by the ponderomotive force of a terawatt (TW), femtosecond (fs) laser pulse  \cite{Dawson-PRL-1979}--\cite{Matlis-NPh-2006}.\ In the non-uniform plasmas, such as a plasma column \cite{Andreev-PoP-2001} produced by optical field ionization (OFI) \cite{Popov-UFN-2004}, or a pre-formed plasma channel \cite{Andreev-PoP-1997}, the laser wake is partly electromagnetic (EM).\ Its rotational current may emit a THz electromagnetic pulse (EMP), bearing information on the plasma stratification \cite{Gorbunov-Frolov-JETP-1996}--\cite{Volchok-PPhCF-2019}.\ The EMP emission pattern, observed in the frequency range from GHz \cite{Englesbe-2018} to THz \cite{Hamster-PRE-1994,Davoine-SciRep-2016}, may bear an imprint of wake currents.

EM footprint of the wake may be understood using a wave conversion theory \cite{Ginzburg_1960}.\ A $p$-polarized EM beam, impinging onto the stratified plasmas, may be partly converted into an electrostatic EPW \cite{Ginzburg_1960}.\ Conversion takes place near the critical surface, where the laser frequency matches the local electron Langmuir frequency, zeroing out the dielectric permittivity.\ Thus, the component of laser electric field, parallel to the density gradient, resonantly drives the  EPW.\ The process may go in reverse: The EPW excited in the stratified plasmas may be converted into EM waves, generating at every point of the density down-ramp an EM wave with a frequency equal to the local Langmuir frequency \cite{Means-PoF-1981,Hinkel-PoF-1993}.\ For this kind of wave transformation to occur, electron velocity in the EPW must have a component \textit{orthogonal} to the density gradient \cite{Gorbunov-Frolov-JETP-1996}.\ This naturally occurs when OFI creates a long-lived plasma column behind a TW laser pulse.\ Electrons oscillating along the column surface have their velocity coupled to the sharp radial density gradient, generating a localized rotational current.\ At every point of the density side-ramp, this current converts the EPW into a radially polarized EM mode with a frequency equal to the local Langmuir frequency.\ Continuum of these modes forms a broadband THz EMP accompanying the wake \cite{Kalmykov_IEEE_2019}.\ The EMP is evanescent in the radial direction, as the subluminal phase velocity of the wake does not permit Cherenkov-type radiation.\ Rapidly evanescing short-wavelength components contribute less to the EMP as the distance from the column surface grows.\ At a few-mm distance, the EMP turns into a single-cycle signal, with a kV/m peak-to-peak electric field variation.\ The linear laser wake in a radially stratified plasma has thus a detectable EM component, overlooked by the conventional wake theory, protruding far into the  plasma-free space, carrying information on the laser wake parameters and plasma stratification.

Section \ref{Sec1} of this Report discusses physical origins of rotational currents in the laser wake, such as the background density gradient, a direct current (DC) flowing through the plasma, and a magnetic field induced by this DC.\ Section \ref{Sec2} derives a complete linear model of the wake in the cylindrically symmetric, non-uniform plasmas.\ The source terms defining generation of a slowly varying magnetic field (second-order in the amplitude of the drive laser pulse) are identified.\ Section \ref{Sec3} concentrates on the case of longitudinally uniform plasma column.\ Section \ref{Sec4} presents the physical case of photoionized helium, shedding light onto the structure of rotational current, using cylindrical particle-in-cell (PIC) simulations via code WAKE \cite{Marques-PoP-1998}.\ Section \ref{Sec5} demonstrates the structure of the EMP Fourier harmonics in the column boundary layer and transformation of the EMP into a single-cycle picosecond signal.\ Section \ref{Concl} summarizes the results.

\section{Physical origins of rotational currents in the laser wake field}
\label{Sec1}

An intense optical pulse propagates in the positive $z$ direction through the inhomogeneous, stationary, electro-neutral plasmas.\ The background electron density, $n_\mathrm{bg}(\mathbf{r})$, defines the spatially non-uniform Langmuir frequency $\omega_\mathrm{pe}(\mathbf{r}) \! = \! (4 \pi e^2 n_\mathrm{bg} / m_e)^{1/2}$ and the wave number $k_\mathrm{p} \! = \! \omega_\mathrm{pe} / c$.\ Here, $-|e|$ and $m_e$ are the electron charge and rest mass, $c$ is the speed of light in vacuum, and $\mathbf{r} \! = \! (x,y,z)$.\ The density $n_{e0}$ at the origin (in the longitudinally uniform system -- along the pulse propagation axis) defines the reference Langmuir frequency $\omega_{\mathrm{pe}0}$ and the wave number $k_{\mathrm{p}0}$.\ The plasma may also carry a DC, $\mathbf{j}_\mathrm{dc} \! = \! e n_\mathrm{bg}(\mathbf{r}) \mathbf{v}_\mathrm{dc}(\mathbf{r})$, which, in turn, induces a non-uniform magnetic field $\mathbf{B}_\mathrm{dc}(\mathbf{r})$.\ The ions (owing to their high inertia) remain almost at rest for a few picoseconds after the passing of the pulse \cite{Gorbunov_Solodov_PoP_2003}.\ Plasma dynamics on this time scale is defined by distributions of electron charge and current density, $\rho \! = \! en_e$ and $\mathbf{j} \! = \! \rho\mathbf{v}$, where $n_e(t,\mathbf{r})$ is the electron number density, and $\mathbf{v}(t,\mathbf{r})$ is the non-relativistic velocity of electron fluid, $ |\mathbf{v} | \! \ll \! c$.\

A small-amplitude EPW in the piece-wise uniform plasmas (such as $\nabla n_\mathrm{bg} \! \equiv \! 0$ within each flat section) neither absorbs nor emits radiation.\ This is because the displacement current in the Amp\`ere's Law cancels the current associated with the density perturbations \cite{Shvets-to-Moloney-PRL-2002,Tikhonchuk-to-Moloney-PRL-2002}.\ Hence, a minor violation of quasi-neutrality, $e\delta n \! = \! e\left(n_e \! - \! n_\mathrm{bg}\right) /n_\mathrm{bg}$, $|\delta n| \! \ll \! 1$, generates no rotational perturbations of electron velocity or current.\ In the non-uniform plasma, on the other hand, even a potential velocity (such as $\nabla \! \times \! \mathbf{v} \!    \equiv \! \mathbf{0}$) becomes a source of the rotational current:  $\nabla \! \times \! \mathbf{j} \! = \!  e\left( \nabla n_\mathrm{bg}\right)\times \mathbf{v} \! \not = \! \mathbf{0}$.\ This takes place, for instance, when the laser pulse propagates through a neutral gas and ionizes it, leaving behind a mm-length, flat-top plasma column.\ The density of plasma constituents drops to zero within a micron-thin shell at the border of the column, while the electrons oscillate along the border.\ Coupling their velocity to the radial density gradient produces the azimuthal rotational current, the source of a radially polarized EMP \cite{Kalmykov_IEEE_2019}.\ For the same reason, the wake driven in a leaky channel -- a pre-formed plasma string  with a density depression on axis \cite{Milchberg-PoP-1996} -- is partly electromagnetic \cite{Andreev-PoP-1997}.

The DC flowing along the channel, $\mathbf{j}_\mathrm{dc} \! = \! \mathbf{e}_z j_\mathrm{dc} \! = \! \mathbf{e}_z \sigma_S E_\mathrm{dc}$, also makes the laser wake field partly rotational, through coupling the longitudinal drift velocity $v_\mathrm{dc}$ to the radial gradient of electron density perturbation, $\nabla_\perp \delta n $.\ Here, $E_\mathrm{dc}$ is the external voltage, $\sigma_S \! \sim \! T_e^{3/2}\lambda_{ei}^{-1}$ is the Spitzer's conductivity of the fully ionized plasma \cite{Spitzer}, and  $ \lambda_{ei} \! \approx \! 24 \! - \! \ln (n_{e0}^{1/2} / T_e)$ is the Coulomb logarithm \cite{NRL} (through the end of this section, the density is in units of cm$^{-3}$, $T_e$ in units of eV, and lengths are in microns.) For the drag from collisions to overpower electric field, avoiding generation of runaway electrons, the drift velocity must be smaller than the electron thermal velocity, $v_\mathrm{dc} \! \ll \! v_{T_e} \! = \! 2.45 \! \times \! 10^{-3} c \, T_e^{1/2}$. Hence, the external voltage must be lower than the Dreicer's field $E_D = 5.6\times 10^{-12} (n_{e0} / T_e)\lambda_{ei}Z$ V/m \cite{Bellan_PoP_2019}.\ Electron drift along the cylindrical channel induces an azimuthal magnetic field, $\mathbf{B}_\mathrm{dc} (r_\perp \le r_\mathrm{ch}) \! \approx \! \mathbf{e}_\phi B_\mathrm{dc}(r_\mathrm{ch}) r_\perp / r_\mathrm{ch}$, where $B_\mathrm{dc}(r_\mathrm{ch}) \! = \! (2\pi\sigma_S r_\mathrm{ch}/c)E_\mathrm{dc}$, and $r_\perp^2 \! = \! x^2 \! + \! y^2$.\ Coupling $\mathbf{B}_\mathrm{dc}$ to the density perturbation in the wake changes the low-frequency canonical vorticity, adding another term to the wake rotational current. The constraint $E_\mathrm{dc} < E_D$ limits the normalized magnetic field, $|e|B_\mathrm{dc}/ (m_e \omega_{\mathrm{pe}0}c) \! < \! \mathcal{B} \! \approx \! 1.15\times 10^{-13} n_{e0}^{1/2} T_e^{1/2} r_\mathrm{ch}$.\ For typical plasma channels, $\mathcal{B}$ is of the order of a few percent, meaning that the DC-induced magnetic field is of the same order of magnitude as the electric field in the linear wake.

\section{Electromagnetic fields in the plasma}
\label{Sec2}

A linear electromagnetic wave in a non-uniform plasma is driven by a rotational current: $\left(\nabla^2 \! - \! c^{-2}\partial^2 / \partial t^2 \right)\tilde{\mathbf{B}} \! = \!  -(4\pi/c)\nabla \times \tilde{\mathbf{j}} \! = \! - 4\pi e (\nabla n_\mathrm{bg})\times \tilde{\mathbf{a}}$ \cite{Ginzburg_1960}.\ The normalized velocity, $\tilde{\mathbf{a}} \! \sim \! \mathcal{O}(\tilde{\mathbf{E}}) \! \sim \! \mathcal{O} (\tilde{\mathbf{B}})$, can be found from a linearized momentum equation, $\partial \tilde{\mathbf{a}}/\partial t \! \approx \! [e/(m_ec)] (\tilde{\mathbf{E}} \! + \!  \tilde{\mathbf{a}}\times\mathbf{B}_\mathrm{dc} \! + \! \hat{\mathbf{v}}_\mathrm{dc}\times\tilde{\mathbf{B}})$, where $\hat{\mathbf{v}}_\mathrm{dc} \! = \! \mathbf{v}_\mathrm{dc}/c$.\ Dropping the last two terms (which is justified for waves with supra-thermal phase velocities, propagating over smooth density gradients) and taking account of the Faraday's Law, one finds that the velocity $\tilde{\mathbf{a}}$ is purely rotational, playing the role of a vector potential: $\nabla\times\tilde{\mathbf{a}} \!  = \! -  e\tilde{\mathbf{B}}/(m_ec^2)$.\ The magnetic field equation thus reduces to
\begin{equation}
\label{eq:equation3a}
(\nabla^2 -\hat{\mathcal{L}}) \tilde{\mathbf{B}} + 4\pi e (\nabla n_\mathrm{bg})\times \tilde{\mathbf{a}} = 0,
\end{equation}
with $\hat{\mathcal{L}} \! = \! c^{-2}\partial^2/\partial t^2 \! + \! k_\mathrm{p}^2(\mathbf{r})$. With the Amp\`ere's Law $\hat{\mathcal{L}}\, \tilde{\mathbf{a}} \! = \! [e / (m_e c^2)] \nabla \! \times \! \tilde{\mathbf{B}}$ in mind, Fourier trans-form recasts the vector product in equation (\ref{eq:equation3a}) as $4\pi e\left(\nabla n_\mathrm{bg}\right) \! \times \! \tilde{\mathbf{a}}(\omega,\mathbf{r}) \! = \! \varepsilon^{-1}\left(\nabla\varepsilon\right) \! \times \! \nabla \! \times \! \tilde{\mathbf{B}}(\omega,\mathbf{r})$, where $\varepsilon (\omega,\mathbf{r}) \! = \!  1  -  \omega_{\mathrm{pe}}^2(\mathbf{r}) / \omega^2 \! = \! 1  -  n_\mathrm{bg}(\mathbf{r}) / n_c(\omega) < 1 $ is the dielectric permittivity, and $n_c(\omega) \! = \! m_e\omega^2/(4\pi e^2)$ is the critical density for the radiation component with a frequency $\omega$. This yields a familiar frequency-domain equation $\left[\nabla^2 + (\omega/c)^2\varepsilon(\omega,\mathbf{r})\right] \tilde{\mathbf{B}}(\omega,\mathbf{r})  +  \varepsilon^{-1} \left(\nabla\varepsilon\right)\times \nabla\times\tilde{\mathbf{B}}(\omega,\mathbf{r}) = 0$ \cite{Ginzburg_1960}.\ The EM mode matches the normal response of the plasma at the critical surface, zeroing out the dielectric permittivity.\ In the vicinity of this singular point, the EM wave may transform into an electrostatic Langmuir mode \cite{Ginzburg_1960} and vice versa \cite{Hinkel-PoF-1993}.\ Radiation of the laser pulse in the underdense plasma, for which $\varepsilon \! \approx \! 1$, cannot undergo this kind of transformation.\ In this physical situation, the mode transformation is a nonlinear effect of a second order in the pulse amplitude.

Solving equation (\ref{eq:equation3a}) for a cylindrically symmetric, parabolic leaky channel \cite{Milchberg-PoP-1996}, $n_\mathrm{bg}(r_\perp \le r_\mathrm{ch}) \! = \! n_{e0}\left(1 \! + \! r_\perp^2 / r_\mathrm{ch}^2\right)$ and $n_\mathrm{bg}(r_\perp \! > \! r_\mathrm{ch}) \equiv 0$, one finds that a Gaussian laser pulse with a waist size $r_0 \! = \! \left( 2r_\mathrm{ch} /k_{\mathrm{p}0}\right)^{1/2} \! \gg \! k_{\mathrm{p}0}^{-1}$ and electric field amplitude (normalized to $m_e \omega_{\mathrm{pe}0}c/|e|$)
\begin{equation}
\label{eq:equation9b}
a(\zeta, r_\perp)  =  a_0\mathrm{e}^{-(r_\perp/r_0)^2} \mathrm{e}^{-\zeta^2/(2L)^2}
\end{equation}
propagates without diffracting, at a slightly sub-luminal group velocity,  $v_{\mathrm{g}0} = c^2 k_0/\omega_0 = \sqrt{\varepsilon_L}c$, $\varepsilon_L \! = \! 1 \! - \! \omega_{\mathrm{pe}0}^2 / \omega_0^2$ (hence $\zeta \! = \! z \! - \! \sqrt{\varepsilon_L}ct $.) Without a channel, the pulse freely diffracts, its amplitude being close to (\ref{eq:equation9b}) in the vicinity of the focal plane, $z \! = \! 0$.\ Considering cylindrically symmetric systems, we describe the wake evolution using co-moving detector variables, $(\zeta, r_\perp, z)$ \cite{Gordon-PoP-2002}, so that $\partial / \partial t \! = \! -  c \sqrt{\varepsilon_L} \partial/\partial\zeta$.\ For all quantities related to the wake, $\partial /\partial z  \approx  \partial /\partial \zeta$, and $\nabla = \mathbf{e}_z \partial /\partial \zeta + \mathbf{e}_\perp \partial /\partial r_\perp$. For the stationary background, $\nabla n_\mathrm{bg} = \left(\mathbf{e}_z \partial /\partial z + \mathbf{e}_\perp \partial /\partial r_\perp\right)n_\mathrm{bg}$.


Ponderomotive force of the pulse (\ref{eq:equation9b}) drives the second-order, low-frequency plasma current, $\mathbf{j}_2 \! = \! e n_\mathrm{bg} c (\hat{\mathbf{v}}_2 \! + \! \delta n \hat{\mathbf{v}}_\mathrm{dc}) \! \sim \! \mathcal{O}(a^2)$, where $\hat{\mathbf{v}}_2 \! = \! \mathbf{v}_2 / c$, and $ \nabla\times\hat{\mathbf{v}}_2 \! = \! - [e/(m_e c^2)]\left(\mathbf{B}_2 \! - \! \delta n \mathbf{B}_\mathrm{dc}\right) $.\ The velocity $\hat{\mathbf{v}}_2 \! = \! \mathbf{a}_2 \! + \! \hat{\mathbf{v}}_S$ is a combination of a magnetic field-driven (normalized) vector potential, $\mathcal{L}_\zeta \mathbf{a}_2 \!  = \! [e/(m_e c^2)] \nabla \! \times \! \mathbf{B}_2$, and a ponderomotively-driven velocity, $\mathcal{L}_\zeta \hat{\mathbf{v}}_S \! = \! (\sqrt{\varepsilon_L}/4) \nabla \partial a^2 / \partial \zeta$, with $\mathcal{L}_\zeta \! = \! \varepsilon_L\left[\partial /\partial\zeta \! - \! \mu(r_\perp, z)\right]^2 \! + \! k_\mathrm{p}^2$.\ The decrement $\mu \! = \!  \mu_0 |\nabla n_\mathrm{bg}| / \mathrm{max}|\nabla n_\mathrm{bg}| \ll k_{\mathrm{p}0}$, with $\mu_0$
defined through first-principle simulations (cf.\ section \ref{Sec4}), accounts for the decay of laser wake fields at a rate proportional to the density gradient, caused by phase mixing of Langmuir oscillations in the density ramp area. The low-frequency magnetic field obeys the driven wave equation,
\begin{equation}
\label{eq:equation9f}
( \nabla^2 - \mathcal{L}_\zeta) \mathbf{B}_2  +   4\pi e (\nabla n_\mathrm{bg})\times \mathbf{a}_2 = - \sum_{j=1}^3 \mathbf{S}_j(\zeta,r_\perp,z).
\end{equation}
Coupling ponderomotively driven velocity to the density gradient defines the source term $\mathbf{S}_1 \! = \!  4\pi e (\nabla n_\mathrm{bg}) \! \times \! \hat{\mathbf{v}}_S$.\ Coupling the DC and its magnetic field to the density perturbation in a wide channel, $k_{\mathrm{p}0}r_\mathrm{ch} \! \sim \! (k_{\mathrm{p}0}r_0)^2 \! \gg \! 1$, yields the terms $\mathbf{S}_2 \! = \! (4\pi/c) \nabla \! \times \! \left(\mathbf{j}_\mathrm{dc} \delta n \right)$ and $\mathbf{S}_3 \! = \! k_\mathrm{p}^2 \, \mathbf{B}_\mathrm{dc}\delta n$.\ The source $\mathbf{S}_1$ is always dominant, $\mathbf{S}_2$ is always a small correction, and $\mathbf{S}_3$ becomes comparable with $\mathbf{S}_1$ if $k_{\mathrm{p}0} r_\mathrm{ch} \! > \! \hat{v}_\mathrm{dc}^{-1/2} \! > \! 20 (T_e[\mathrm{eV}])^{-1/4}$.\ The cylindrically symmetric driver (\ref{eq:equation9b}) does not impart an azimuthal component to $\hat{\mathbf{v}}_S$.\ The source is thus azimuthally polarized, $\mathbf{S}_j \! = \! \mathbf{e}_\phi S_{j\phi}(\zeta,r_\perp)$, and so is the low-frequency magnetic field.\ Here, $S_{1\phi} \! = \! -4\pi e [(\partial n_\mathrm{bg} / \partial r_\perp) \hat{v}_{Sz} \! - \! (\partial n_\mathrm{bg} / \partial z)\hat{v}_{S\perp}]$, $S_{2\phi} \! \approx \! - (4\pi/c)( \partial \delta n /\partial r_\perp ) j_\mathrm{dc} $, and $S_{3\phi} \! = \! k_\mathrm{p}^2\, B_\mathrm{dc}\delta n$.\

\section{EMP from rotational wake current in photoionized plasma column: $\mathbf{S}_2 \! = \! \mathbf{S}_3 \! \equiv \! \mathbf{0}$}
\label{Sec3}

Laser wake fields in a longitudinally uniform, flat-top plasma column, in the vicinity of the laser focus, $z \! = \! 0$, depend solely on $\zeta$ and $r_\perp$, with the source $S_{1\phi}(\zeta,r_\perp) \! = \! -4\pi e (\mathrm{d} n_\mathrm{bg} / \mathrm{d} r_\perp) \hat{v}_{Sz}$, and $\mathcal{L}_\zeta  \hat{v}_{Sz} \! = \! (\sqrt{\varepsilon_L}/4) \partial^2 a^2 / \partial \zeta^2$.\ With all lengths normalized to $k_{\mathrm{p}0}^{-1}$, the Amp\`{e}re's Law reduces to $\hat{\mathcal{L}}_\zeta a_{2z} \! = \! r_\perp^{-1}\partial (r_\perp \hat{B}_{2\phi}) /\partial r_\perp$, and equation (\ref{eq:equation9f}) to
\begin{eqnarray}
\nonumber
(\partial /\partial r_\perp)\, r_\perp^{-1}\partial(r_\perp \hat{B}_{2\phi})/\partial r_\perp + ( \partial^2 / \partial \zeta^2 - \hat{\mathcal{L}}_\zeta ) \hat{B}_{2\phi} & - & (\mathrm{d} n / \mathrm{d} r_\perp) \hat{a}_{2z} \\
 & = & (\mathrm{d} n / \mathrm{d} r_\perp) \hat{v}_{Sz},\label{eq:equation1a}
\end{eqnarray}
Here, $\hat{\mathcal{L}}_\zeta \! = \! k_{\mathrm{p}0}^{-2} \mathcal{L}_\zeta$, $n \! = \! n_{\mathrm{bg}}/n_{e0}$, and $\hat{B}_{2\phi} \! = \! |e| B_{2\phi} / (m_e \omega_{\mathrm{pe}0}c)$.\ Fourier image of the magnetic field, $B_{2\phi}^k (r_\perp) \! = \! \int \hat{B}_{2\phi}(\zeta, r_\perp) \mathrm{e}^{\mathrm{i}k\zeta}\,\mathrm{d}\zeta$, expresses through the image of the drive pulse intensity, $a_k^2 = \sqrt{2\pi} a^2 L  \mathrm{e}^{-(kL)^2 / 2} \mathrm{e}^{-2(r_\perp/r_0)^2}$ and the dielectric function $\varepsilon_k$:
\begin{eqnarray}
&& (\mathrm{d} / \mathrm{d} r_\perp + k^{-2}W)\, r_\perp^{-1}\mathrm{d} (r_\perp B_{2\phi}^k) / \mathrm{d} r_\perp -  k^2(1 - \varepsilon_k \varepsilon_L)B_{2\phi}^k = S_k, \label{eq:equation2a} \\
\label{eq:equation2b}
&& \varepsilon_k(\mu, r_\perp) =  \left[ 1 + \mathrm{i}\mu(r_\perp)/k \right]^2  -  n(r_\perp)/(\varepsilon_L k^2).
\end{eqnarray}
Here, $S_k \! = \! (\sqrt{\varepsilon_L}/4) W a_k^2$ is the Fourier image of the source, and $W \! = \! \left( \varepsilon_k \varepsilon_L \right)^{-1}\mathrm{d} n / \mathrm{d} r_\perp$.\ $B_{2\phi}^k$ and $a_k^2$ define Fourier images of the electric field components: $ E_{2z}^k \!  = \!  \mathrm{i} (k \varepsilon_k \varepsilon_L )^{-1} [ \sqrt{\varepsilon_L}  ( 1 \! + \! \mathrm{i}\mu / k)^2 \, r_\perp^{-1}\mathrm{d}(r_\perp B_{2\phi}^k) /\mathrm{d} r_\perp \! - \! na_k^2/4]$, and $ E_{2r}^k \! = \! ( k^2\varepsilon_k \varepsilon_L )^{-1} [k^2 \sqrt{\varepsilon_L} \, ( 1 \! + \! \mathrm{i}\mu / k )^2 B_{2\phi}^k \! + \! ( r_\perp/r_0^2 ) \, na_k^2]$.\  As $n(r_\perp \! \to \! \infty) \! \to \! 0$, contributions from the ponderomotive force vanish, making electric field in the plasma-free area purely rotational.\ Potential electric field of the Langmuir wave transforms into the rotational field of a vacuum eigenmode near the critical surface, a cylinder with the radius $R_\mathrm{crit}(k)$ found from the transcendental equation $\varepsilon_k(0,R_\mathrm{crit}) \! = \! 0$ for every wave number from the band $0 \! < \! k \! < \! 1/\sqrt{ \varepsilon_L}$.\ The EMP thus contains the entire frequency band $0 \! < \! \nu \! < \! \nu_0 \! = \! \omega_{\mathrm{pe}0}/(2\pi)$, where $\nu \! = \! k\omega_\mathrm{pe}/(2\pi)$.\ Attenuation of the wake keeps $E_{2z}^k(R_\mathrm{crit})$, $ E_{2r}^k(R_\mathrm{crit})$, and the source image $S_k(R_\mathrm{crit})$ bounded.

The magnetic field, proportional to the source, vanishes at the axis: $B_{2\phi}^k (0) \! = \! \mathrm{d} B_{2\phi}^k / \mathrm{d} r_\perp (0) \! = \! 0$.\ The non-trivial asymptotic bound at infinity is proportional to the modified Bessel function of the second kind, $B_{2\phi}^k (r_\perp \! \to \!  \infty) \! \sim \!  CK_1( U r_\perp)$, and $\mathrm{d} B_{2\phi}^k / \mathrm{d} r_\perp (r_\perp \! \to \! \infty) \! \sim \!  - C (U/2) [ K_0( U r_\perp ) \! + \! K_2( U r_\perp )]$, where $U \! = \!  k r_\perp \sqrt{1 \! - \! \varepsilon_L}$.\ Matching solutions within and without the column at $r_\perp \! = \! R_\mathrm{crit}$  eliminates the constant $C$ and defines the phase of the solution at infinity.

The source propagates at a subluminal group velocity of the laser pulse, $v_{\mathrm{g}0} \! = \! c\sqrt{\varepsilon_L}$.\ The magnetic field thus exponentially evanesces for $r_\perp \! > \! U^{-1}$: $B_{2\phi} ^k \! \sim \! \sqrt{\pi/(2 Ur_\perp)} \exp(-Ur_\perp)$ \cite{Gradshteyn-Ryzhik}.\ This ``detection'' zone is a few millimeters away from the column, safe to place a detector, e.g.\ an electro-optical crystal \cite{Curcio-PRAppl-2018}, avoiding its exposure to the stray optical radiation or a gas flow of the target.\ Rapid decay of short-wavelength components makes the components with $k \! \ll \! 1$ dominant in the detection zone, transforming the THz signal into a single-cycle picosecond EMP.

\begin{figure}[t]
\centering
\includegraphics[scale=0.825]{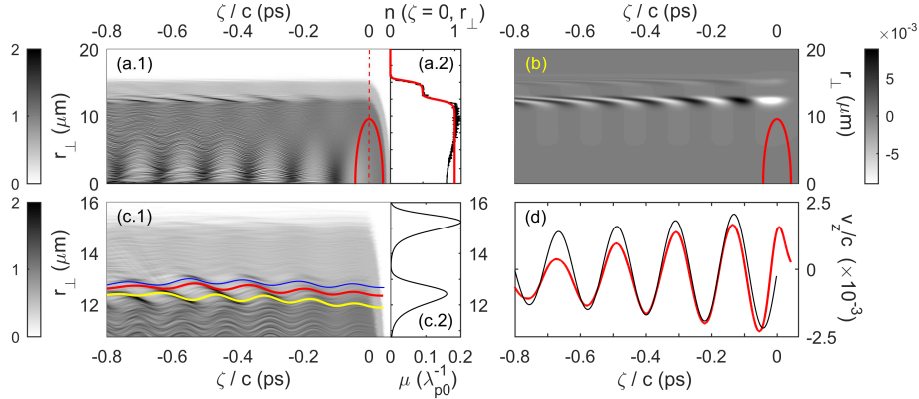}
\caption{\label{Figure1} (a) OFI of helium leaves behind the drive pulse (propagating to the right) a flat-top plasma column (WAKE simulation.) Red iso-contour of intensity (at $\mathrm{e}^{-2}$ of the peak) marks the pulse position [also in panel (b).] (a.2) Density ramps in the boundary layer, together with their analytic fit (red.) This fit is used in analytically calculating the EMP source (b), the local decrement of wake attenuation (c.2), and the electron velocity [panel (d), black.] (c.1) The side-ramp fragment from panel (a).\ Superimposed trajectories of three sample macroparticles show the onset of phase mixing in finite-amplitude radial oscillations.\ These oscillations decay with a local decrement depicted in (c.2).\ The decrement peaks near the mid-point of each  ramp, and vanishes within flat areas.\ (d) Red: Damped oscillations in longitudinal velocity of a WAKE macroparticle [the ``red'' trajectory from (c.1), the average radial offset $R_\mathrm{c} \approx 12.65$ $\mu$m.] Black: $\hat{v}_2 \! = \! a_{2z} \!  + \! \hat{v}_{Sz}$ obtained semi-analytically for $r_\perp \! = \!  R_c$, using the decrement from (c.2).}
\end{figure}

\section{Physical case of photoionized helium column}
\label{Sec4}

The pulse with an envelope (\ref{eq:equation9b}) and a carrier wavelength $\lambda_0 \! = \! 0.8$ $\mu$m ($\omega_0 \! = \! 2\pi c/\lambda_0 \! = \! 2.356 \! \times \!  10^{15}$ s$^{-1}$) focuses at $z \! = \! 0$.\ Its full width at half-maximum in intensity is $\tau_L \! = \! \sqrt{8\ln2}L/c \! = \! 50$ fs; $r_0 \! = \! 9.536$ $\mu$m; and $a_0 \! = \! 0.5$.\ The pulse average power is $P \! \approx \! 0.815$ TW, total energy $E \! \approx \! \tau_L P \! \approx \! 41$ mJ.\ The  peak intensity $I_\mathrm{peak} \! = \! 5.4 \! \times \! 10^{17}$ W/cm$^2$ is sufficient to fully strip helium, generating the background electron density $n_{e0} \! = \! 7 \! \times \! 10^{17}$ cm$^{-3}$.\ This defines the Langmuir frequency $\omega_{\mathrm{pe}0} \! = \! 4.72 \! \times \! 10^{13}$ s$^{-1}$ (the oscillation period $ \nu_0^{-1} \! \approx \! 0.133$ ps), and $k_{\mathrm{p}0}^{-1} \! = \! c/\omega_{\mathrm{pe}0} \! \approx \! 6.36$ $\mu$m (Langmuir wavelength  $\lambda_{\mathrm{p}0} \! = \! 2\pi/k_{\mathrm{p}0} \! = \! 40$ $\mu$m.) Hence, $k_{\mathrm{p}0}L \! = \! 1$ (the resonant condition for wake excitation), $k_{\mathrm{p}0}r_0 \! = \! 1.5$, and  $\varepsilon_L \! = \! 0.9996$.\ A first-principles simulation via quasistatic, cylindrical PIC code WAKE \cite{Marques-PoP-1998} sheds light onto the wake dynamics at the border of the column.\ The code uses co-moving detector variables $(\xi, r_\perp, z)$ with $\xi \! = \! z \! - \! ct$ (the pulse propagating in the positive $z$ direction.) The correct pulse group velocity is retrieved by using an  extended paraxial operator.\ OFI of neutrals is described in terms of Keldysh model with a linearly polarized pulse \cite{Marques-PoP-1998}.\ Ions, even though not frozen, remain almost at rest due to their high inertia.\ The newly-born free electrons are pushed by the optical cycle-averaged ponderomotive force.\ The resulting charge separation and electron currents (varying slowly over the optical cycle) form fully electromagnetic, cylindrical plasma wake.\ Synchronization of macroparticles with the wake (longitudinal wavebreaking \cite{Akhiezer-JETP-1956}) is forbidden by the quasistatic approximation; yet the trajectory crossing in transverse oscillations is permitted \cite{Marques-PoP-1998}.




Figure \ref{Figure1}(a.1) displays the plasma column in the vicinity of the pulse focal plane.\ A stair-like electron density profile in figure \ref{Figure1}(a.2) reflects two ionization stages of fully-stripped helium.\  The density falls off within two micron-scale down-ramps separated by a narrow shelf.\ This profile is approximated analytically and fitted into equation (\ref{eq:equation1a}).\ The semi-analytically defined source is displayed in figure \ref{Figure1}(b).\ It is localized within two narrow concentric shells at the location of the density ramps.\ Velocities contributing to the source are non-relativistic, $| \hat{v}_{Sz} | \! \sim \!  10^{-3}$.\


\begin{figure}[t]
\centering
\includegraphics[scale=0.9]{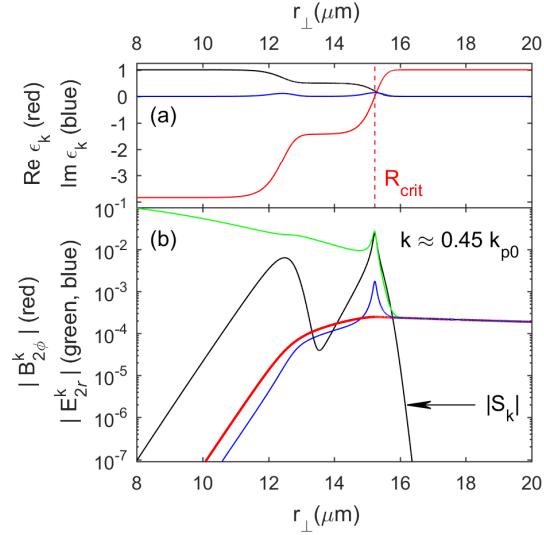}
\caption{\label{Figure2} (a) Stair-like radial profile of the background electron density (black).\ Real part of the dielectric permittivity (\ref{eq:equation2b}) (red) changes sign at the critical surface, while $\mathrm{Im}\,\varepsilon_k$ (blue) is small yet non-vanishing.\ (b) Fourier image of the source $S_k$ (black) has maxima near a mid-point of the high-density ramp and at the critical surface.\  The radial electric field $E_{2r}^k$, both full (green) and magnetic field-driven (blue), sharply peaks at the critical surface.\ Red: The magnetic field (\ref{eq:equation2a}) varies smoothly, slowly evanescing in the plasma-free space and falling off sharply towards the plasma interior. }
\end{figure}

Figure \ref{Figure1}(c.1) shows that the electrons situated further away from axis sample lower densities, and thus oscillate with longer periods.\ The phase shift between neighboring fluid elements (WAKE macroparticles) accumulates with time, gradually increasing phase front curvature of the wave.\ Any pair of trajectories, separated by less than the amplitude of their radial oscillations, will cross within a finite period of time.\ This phase mixing \cite{Marques-PoP-1998,Andreev-PoP-2001} eventually sets in for all particles oscillating in the ramp area, destroying the wake \cite{Pukhov-PRL-1997}.\ In our quasi-linear regime, phase mixing is not catastrophic, causing slow exponential decay in macroparticle velocity oscillations [cf.\ figure \ref{Figure1}(d).] Figure \ref{Figure1}(c.2) shows the local decay rate, $\mu \! = \! \mu_0 |\mathrm{d} n / \mathrm{d} r_\perp|/\mathrm{max}|\mathrm{d} n / \mathrm{d} r_\perp|$, obtained using the analytic fit $n(r_\perp)$ from figure \ref{Figure1}(a.2), with $\mu_0 = k_{\mathrm{p}0}/(10\pi)$ assessed from WAKE macroparticle tracking; this dependence is used in numerically solving equation (\ref{eq:equation2a}).

\section{Single-cycle EMP accompanying the laser wake}
\label{Sec5}

Equation (\ref{eq:equation2a}) is solved  for the physical parameters of section \ref{Sec4}.\ Figure \ref{Figure2}(a) shows the structure of complex dielectric permittivity (\ref{eq:equation2b}).\ The wave number corresponds to the critical surface laying at the mid-point of the low-density ramp, $R_{\mathrm{crit}} = 15.2$ $\mu$m.\ Fourier images of the source and of the radial electric field in figure \ref{Figure2}(b) sharply peak at the critical surface.\ Fourier component of the source has another maximum at the mid-point of the high-density ramp, yet wave transformation at the the critical surface contributes to the EMP the most.\ Rotational fields making up the EMP slowly evanesce in the plasma-free space.\

\begin{figure}[t]
\centering
\includegraphics[scale=0.85]{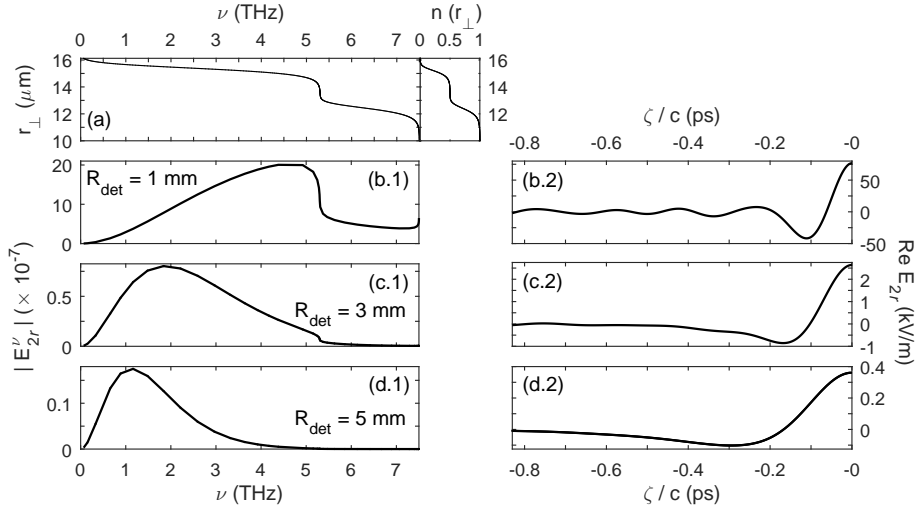}
\caption{\label{Figure3} THz signal in the detection zone.\ As the distance from the column increases, the higher-density areas contribute less to the signal.\ Stronger evanescence of short-wavelength components results in formation of a single-cycle THz signal with a kV/m-scale electric field.\ (a) Radiation frequency $\nu = k[\omega_{\mathrm{pe}0}/(2\pi)]$ maps onto the density profile (inset.) There is a single-valued correlation between the frequency and the radial offset at which radiation with this frequency originates.\  (b.1)--(d.1) As the distance from the column increases, the peaks in frequency spectra shift towards $\nu \approx 1$ THz, the shape of the spectra losing connection to the structure of the column boundary. (b.2)--(d.2) The single-cycle, half-ps EMP forms, with a 0.4--4 kV/m peak-to-peak electric field variation.}
\end{figure}

In the detection zone, the EMP is radially polarized, the amplitudes of radial electric and azimuthal magnetic fields almost equal to each other.\ Weakly evanescent low-frequency spectral components, generated at the very foot of the low-density ramp, define the shape of the signal.\ Frequency spectra in figures \ref{Figure3}(b.1)--(d.1) show progressive red-shift as the detector moves from $R_\mathrm{det} \! = \! 1$ mm to 5 mm, the average frequency drifting from 4.2 to 1 THz.\ Close to the near-field zone [$R_\mathrm{det} \! = \! 1$ mm, figure \ref{Figure3}(a)], the spectrum bears an imprint of the boundary layer structure.\ The part of the spectrum generated at the high-density ramp makes a low-amplitude shelf at higher frequencies, $\nu \! > \! 5.3 $ THz.\ Deep in the detection zone [figures \ref{Figure3}(c.1) and \ref{Figure3}(d.1)], the higher-frequency part of the spectrum is suppressed, the shape of the spectrum losing connection to the structure of the boundary layer.\ As the spectral peak shifts towards 1 THz, the time-domain signal reshapes into a single-cycle picosecond pulse with a kV/m peak-to-peak voltage variation.\ Qualitatively, this evanescent EMP replicates the waveform of the source and the plasma wake seen in figures \ref{Figure1}(a.1) and \ref{Figure1}(b), the high-frequency content concentrated closer to the rim of the column, the low-frequency content spacing further away.\ The pulsed electric field of the EMP, detectable via electro-optical sampling \cite{Curcio-PRAppl-2018} or with a small antenna -- a mm-length conductive pin -- set about the path of the drive laser pulse, indicates the presence of the wake, helping assess the efficiency of pulse coupling to the photoionized plasma.

\section{Summary}
\label{Concl}


Laser wake in the stratified plasmas has electromagnetic signatures, detectable at a macroscopic distance in the plasma-free area. Particularly, the surface of a cylindrical plasma column, created by OFI of a neutral gas, supports an azimuthally polarized rotational current, the source of radially polarized evanescent EMP.\ This current is produced by coupling the ponderomotively-driven, longitudinal oscillations in electron velocity to the radial gradient of electron density.\ Radial evanescence of the EMP is dictated by the subluminal phase velocity of the plasma wake.\ A physical demonstration with a 50 fs-long, sub-TW laser pulse shows the presence of a kV/m-level, radial electric field at a few-mm distance from the column.\ This is  possibly detectable via electro-optical sampling~\cite{Curcio-PRAppl-2018}.\ The field of the wake-driven THz EMP is composed of the continuum of frequency components in the band $0 \! < \! \omega \! < \! \omega_{\mathrm{pe}0}$, generated at the column border.\ As the observer moves further away in the radial direction, the high-frequency components, $\omega \! \approx \! \omega_{\mathrm{pe}0}$, become suppressed.\ This shifts the EMP spectrum toward longer wavelengths, eventually forming a single-cycle, radially polarized picosecond pulse at a mm-scale distance.\ Detecting and spectrally resolving the THz signal should provide information on the life-time of the plasma wake, the background electron density in the column, and the intensity of pulse creating the column and driving the wake.


\ack The work is supported by the AFRL contract FA9451-17-F-0011, and AFOSR grants FA9550-16RDCOR325, FA9550-19RDCOR023, and FA9550-19RDCOR027.

\end{document}